\documentclass[aps,prb,twocolumn,groupedaddress,showpacs,floatfix,altaffilletter]{revtex4-1}
\usepackage[utf8]{inputenc}    
\usepackage[T1]{fontenc}
\usepackage{graphicx}
\usepackage{grffile}
\usepackage{amsmath}
\usepackage{amssymb}
\usepackage{bm}
\usepackage{color}
\usepackage[dvipsnames]{xcolor}
\usepackage{amsmath,amssymb,amsfonts}
\usepackage{epsfig}
\usepackage{times}
\usepackage{float}
\usepackage{gensymb}
\usepackage[colorlinks,bookmarks=false,citecolor=blue,linkcolor=red,urlcolor=blue]{hyperref}

\newcommand{\fref}[1]{Fig.~\ref{#1}}
\newcommand{\eref}[1]{Eq.~(\ref{#1})}
\newcommand{\sref}[1]{Sec.~(\ref{#1})}
\newcommand{\tref}[1]{Table~\ref{#1}}

\begin{document}

\title{Electronic and Magnetic Properties of double-perovskites La$_2$MnRuO$_6$ and LaA''MnFeO$_6$ (A'' = Ba, Sr, Ca) and their Potential for Magnetic Refrigeration }

\author{C. Gauvin-Ndiaye$^{1}$,  A.-M.S. Tremblay$^{1,2}$,  and R.~Nourafkan$^{1}$}
\affiliation{$^1$Institut quantique, D{\'e}partement de Physique and RQMP, Universit{\'e} de Sherbrooke, Sherbrooke, Qu{\'e}bec, Canada  J1K 2R1}
\affiliation{$^2$Canadian Institute for Advanced Research, Toronto, Ontario, Canada M5G 1Z8}
\date{\today}
\begin{abstract}
Magnetic refrigeration at room-temperature is a technology that could potentially be more environmentally-friendly, efficient and affordable than traditional refrigeration. The search for suitable materials for magnetocaloric refrigeration led to the study of double-perovskites La$_2$MnNiO$_6$, La$_2$MnCoO$_6$ and La$_2$MnFeO$_6$. While La$_2$MnNiO$_6$ and La$_2$MnCoO$_6$ are ferromagnets with near room-temperature $T_C$s, previous theoretical study of double-perovskite La$_2$MnFeO$_6$ revealed that this material is a ferrimagnet due to strong electronic interactions in Fe-$d$ orbitals. Here, we investigate the double-perovskites La$_2$MnRuO$_6$ and LaA''MnFeO$_6$ (A'' = Ba, Ca and Sr) with density functional theory (DFT) as materials that can counteract the effects the strong repulsion present in the in Fe-$d$ shells of La$_2$MnFeO$_6$ and lead to a ferromagnetic state. Our study reaveals that while La$_2$MnRuO$_6$ is also a ferrimagnet, but with a higher net magnetic moment per formula than La$_2$MnFeO$_6$, doubly-ordered LaA''MnFeO$_6$ are ferromagnets. By mapping the total energy of the LaA''MnFeO$_6$ compounds obtained from DFT calculations to the Ising model, we also calculate their magnetic exchange couplings. This allows us to estimate the trend in $T_C$ of the three doped La$_2$MnFeO$_6$ materials with classical Monte-Carlo calculations and predict that doubly-ordered LaBaMnFeO$_6$ and LaSrMnFeO$_6$ could be suitable materials for room-temperature magnetic refrigeration.
\end{abstract}
    
\maketitle
\section{Introduction}
Traditional cooling utilizes refrigerant gases that are harmful to the environment due to their global warming potential. \cite{ref-refri} A promising alternative is magnetic refrigeration (MR). It relies on the magnetocaloric effect, which describes the variation of the temperature of a material subjected to a change in magnetic field. 
The magnetocaloric materials are generally ferromagnets that undergo a phase transition to a paramagnetic state around operating temperature. The magnetocaloric materials suitable for domestic refrigeration have a transition temperature around room temperature.  
    
The design of new magnetocaloric materials is one of the main research areas in the field. 
    %
Double-perovskites La$_2$MnNiO$_6$ (LMNO) and La$_2$MnCoO$_6$ (LMCO) have been proposed and investigated extensively because they are ferromagnetic insulators with large total moments, $5$ and $6 \mu_B$/f.u. respectively~\cite{ref-lmno, ref-lmco}. Furthermore, they are low-cost, resistant to corrosion and recyclable compounds. 
Their Curie temperatures $T_C$ are however below room temperature, respectively $280$ K and $226$ K~\cite{ref-lmno,ref-lmco226}.  In order to be able to use these double-perovskite oxides for MR, one would have to find a way to increase their $T_C$s. 
The natural assumption is that a similar compound, La$_2$MnFeO$_6$ (LMFO), is also a ferromagnetic insulator, but with a higher $T_C$ due to a possibly higher magnetic moment on Fe. However, experiments show that LMFO is a ferrimagnet with anti-parallel alignment of Mn and Fe moments on neighboring sites~\cite{PhysRevB.91.054421}. Further theoretical investigation has shown that in LMNO and LMCO, the Mn ions acquire Mn$^{4+}$ oxidation state with three electrons in their $t_{2g}$ orbitals, leaving a doubly degenerate Mn-$e_g$ to contribute in the superexchange mechanism with O-$p$ orbitals. The almost unoccupied, doubly degenerate Mn-$e_g$ sets the stage for Hund's coupling to be effective by reducing the ferromagnetic (FM) ground-state energy in comparison with the antiferromagnetic (AFM) one~\cite{ref-ourlmnopaper}. However, in LMFO, large electronic correlations prevent double occupancy in Fe $d$-shells, promoting the Fe$^{3+}$ valence state with half-filled $d$-shell over Fe$^{2+}$, and leading to high-spin Mn$^{3+}$ and Fe$^{3+}$ states~\cite{ref-ourlmnopaper}. Mn$^{3+}$ has four valence electrons with one residing on $e_g$ states that lifts their degeneracy due to the Jahn-Teller mechanism. Therefore, in LMFO, the effective shells of the two transition metal ions are half-filled Mn-$e_g$ and Fe-$d$, leading to the usual antiferromagnetic superexchange interaction. Because the magnitude of the down-spin magnetic moment is different from the up-spin magnetic moment, the resulting state is a ferrimagnet. We use antiferromagnet (AFM) and ferrimagnet interchangeably since the magnetic moments of the two magnetic ions are different, but we will mostly use the acronym AFM. 
    
The main factor that drives LMFO to be a ferrimagnet is the strong electron-electron interaction in Fe-$d$ shells that overcomes the crystal field splitting in Mn-$d$ shells. 
In order to promote Mn$^{4+}$ oxidation state and to design materials that could potentially be more suitable than LMNO and LMCO for magnetic refrigeration, we consider and discuss two solutions. First, we study the double-perovskite La$_2$MnRuO$_6$ (LMRO). This compound is obtained by iso-electronic substitution of Fe with Ru. The valence orbitals of Ru are $4d$. These orbitals are more extended in space than the Fe-$3d$, and hence show less significant electronic correlations. One could then hope for a Mn$^{4+}$-O-Ru$^{2+}$ ferromagnetic superexchange interaction. Second, we consider the effect of hole-doping LMFO by substituting half of the La atoms with A'' = Ba, Sr or Ca, leading to LaA''MnFeO$_6$ with ferromagnetic Mn$^{4+}$-O-Fe$^{3+}$ superexchange interaction in the ordered materials. 
    
In this paper, we employ the real material calculation described in Sec.~\ref{section:method} to find in Sec.~\ref{section:LMRO_properties} the magnetic ground state of LMRO and in Sec.~\ref{section:LAMFO} of  hole-doped LMFO. The latter section contains information on structure optimization, magnetic and electronic ground state properties, estimates of magnetic exchange couplings and of the Curie temperature.

\section{Method} \label{section:method}
We investigate the ground state structural, electronic, and magnetic properties of double-perovskites LMRO, LBMFO, LSMFO and LCMFO with density functional theory calculations. The calculations are performed within the full-potential all electron basis set as implemented in the WIEN2k package, using the PBE GGA functional.~\cite{ref-wien2k, PhysRevLett.77.3865} The interaction effects are taken into account using GGA+U. 
The GGA+U calculations are carried using the approximate correction for self-interaction correction (SIC) as described in Ref.~\onlinecite{PhysRevB.48.16929}.  
In DFT calculations, we check the convergence with respect to the number of $\bf k$-points used in the Brillouin zone and the plane-wave cut-off K\textsubscript{max}, controlled by the parameter R\textsubscript{mt}$\cdot$K\textsubscript{max}, where R\textsubscript{mt} is the muffin-tin radius. Both volume and internal coordinates are fully relaxed. 
We consider several collinear magnetic orderings to obtain exchange coupling between transition metal ions. Consequently, these couplings are used to estimate Curie temperatures using mean-field calculation and Monte Carlo simulations with the GT-GPU method on a cubic lattice.\cite{2018arXiv180109379B}

%

\section{Electronic and magnetic properties of L\MakeLowercase{a}$_2$M\MakeLowercase{n}R\MakeLowercase{u}O$_6$}
\label{section:LMRO_properties}
We first studied LMRO in order to see if a ferromagnetic superexchange interaction is possible between Mn and Ru ions, considering the fact that electronic repulsion in Ru-4$d$ shells are less important than in Fe-$3d$ shells. We used experimental crystal structure data to perform the calculations, without structure relaxation~\cite{ref-lmro}. This experiment found LMRO to be a ferrimagnet. However, it was carried on disordered LMRO with space group $Pbnm$. We added a rock-salt ordering for Mn and Ru atoms, which lowers the space group symmetry to $P2_1/c$. The same space group has been observed experimentally in similar A$_2$B'B''O$_6$ compounds with B-site rock-salt order, such as ordered single crystal LMNO and LMCO.\cite{ref-lmno,ref-lmco} In these ordered double-perovskites, B' and B'' atoms alternate in each spatial direction. Here, we are interested in seeing if the addition of B-site order in LMRO can drive a ferromagnetic ground state. 

\begin{figure}
    \centering
    \includegraphics[width=0.75\columnwidth]{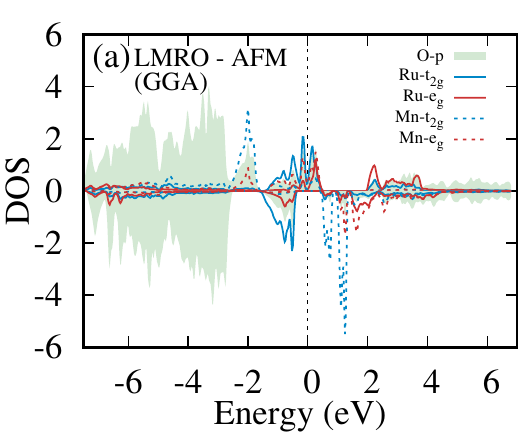} 
    \includegraphics[width=0.75\columnwidth]{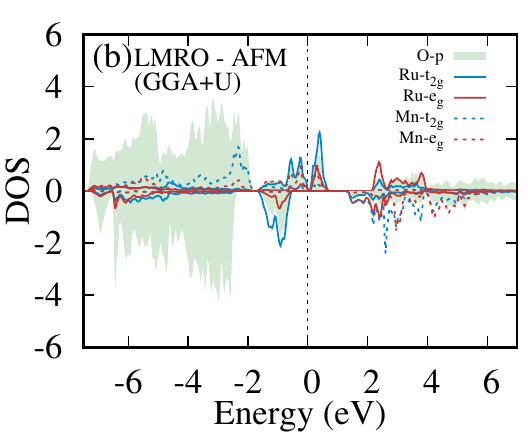}\\ 
    \vspace{0.5cm}
    \includegraphics[width=0.75\columnwidth]{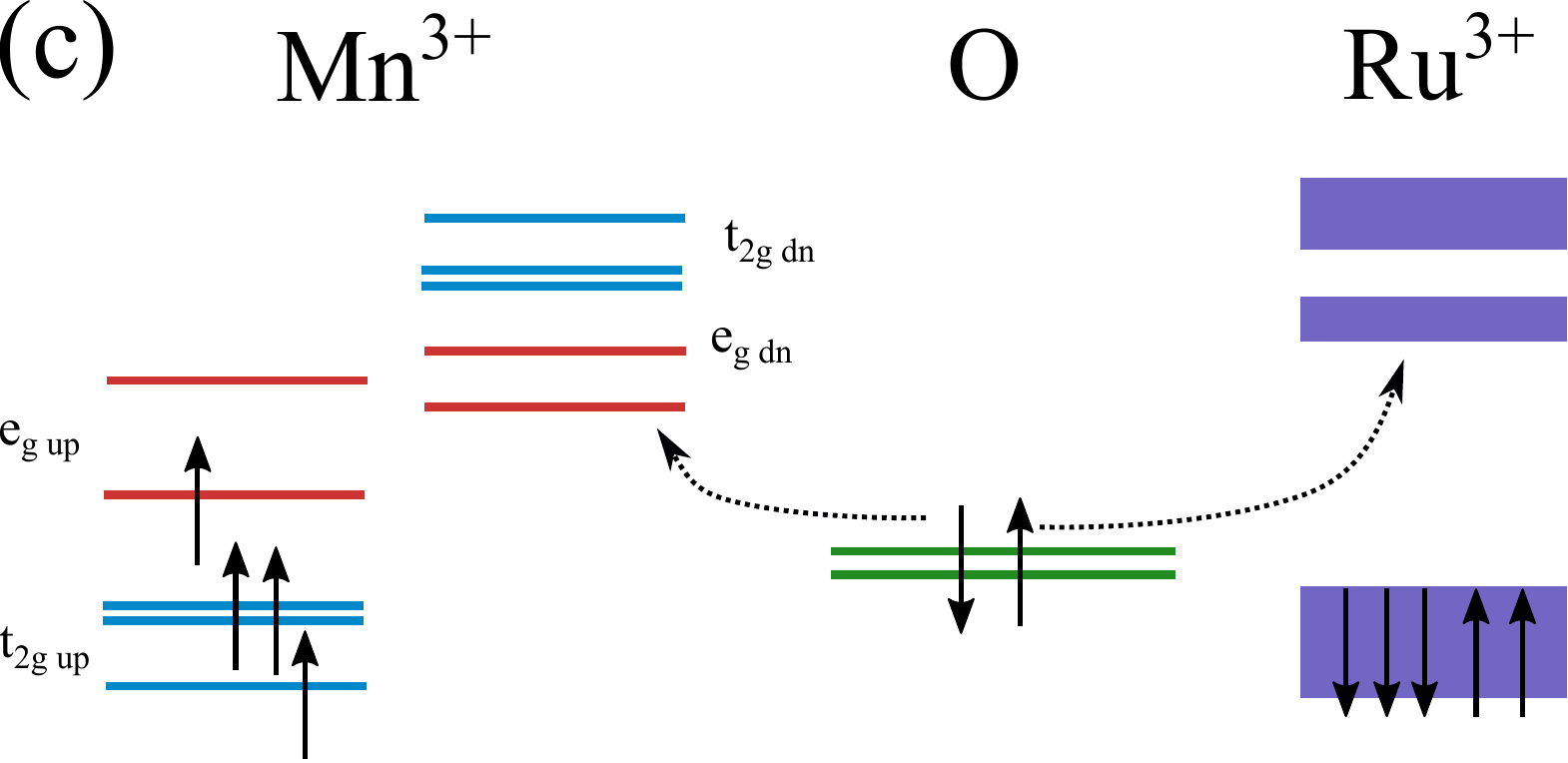}
    \caption{(Color online) Spin-resolved partial density of states for Mn-$e_g$, Mn-$t_2g$, Ru-$e_g$, Ru-$t_{2g}$ and O-$p$ from (a) GGA calculations and (b) GGA+U calculations. GGA+U calculations are performed with $U_{eff}=3$ eV for Mn-$d$ orbitals and $U_{eff}=1.09$ eV for Ru-$d$ orbitals. The upper part in each panel is majority-spin DOS result, and the lower the minority-spin
one. GGA(+U) calculations predict a metallic (insulating) ferrimagnetic ground state for LMRO.
    (c) Schematic representation of the superexchange interaction in ferrimagnetic LMRO. The figure represents schematically the weight of each orbital with respect to the others and is derived from the partial DOS plots and partial charge data.}
    \label{fig:lmro}
\end{figure}

Our GGA calculations show that the ground sate of LMRO is a spin density wave with anti-parallel magnetic moments of the neighboring Mn and Ru ions. \fref{fig:lmro} (a) illustrates the GGA density of states (DOS) of LMRO. As one can see from the figure, the system is in a metallic state with finite spectral weight at the Fermi level. The total moment is $3\mu_B$/f.u. 
Adding the correlation effects within the GGA+U framework opens up a charge gap at the Fermi level leading to an insulating ferrimagnetic ground state. We used $U_{eff}=1.09$ eV for Ru-$d$ shells and $U_{eff} = 3.0$ eV for Mn-$d$ shells. \fref{fig:lmro} (b) shows the GGA+U DOS of LMRO. The band gap is $\simeq 0.03$ eV. Using a larger interaction value for Ru, i.e. $U_{eff} = 3.0$ eV, does not change the ground state magnetic alignment, but increases the band gap to $\simeq 0.2$ eV as expected. As seen in \fref{fig:lmro} (b), the Ru-$d$ orbitals are relatively delocalized, with wide partial DOS, and there is sizable overlap between majority-spin O-$p$ and Ru-$d$ partial DOS. Although the ground state of LMRO is ferrimagnetic, it has a larger net moment than LMFO because the Ru$^{3+}$ ions are in low-spin configuration, and hence have a smaller moment than high-spin Fe$^{3+}$.

In order to confirm the results, we relaxed the structure by optimizing its internal degrees of freedom, without changing the volume of the unit cell. We used $U_{eff}=3$ eV in Mn- and $U_{eff}=1.09$ eV in Ru-$d$ shells to relax the structure. The ground state predicted from the relaxed structures is also ferrimagnetic. This did not change drastically the predicted partial moments, total moment per formula unit or the partial charges.

An ionic picture for Ru ions is not quite applicable. Indeed, in contrast to Mn-$d$ orbitals, in which spin-dependent orbital occupation is almost zero or one and does not depend sensitively on the magnetic order, some of Ru-$d$ orbitals are partially occupied (see \tref{tab:n}). Nevertheless, the picture of localized spins to describe the superexchange mechanism in LMRO works rather well as we now show. In both FM and AFM magnetic configurations, the Mn$^{3+}$ ions are in high-spin configuration. Furthermore, the Mn-$e_g$ degeneracy is lifted due to the Jahn-Teller distortion, hence, only one Mn-$e_g$  orbital is contributing in the superexchange interaction.  Let us assume that Mn majority spin species is up as shown in \fref{fig:lmro} (c). In this case, a down-spin electron of O-$p$ contributes to the superexchange mechanism with Mn-$e_g$ due to Pauli's exclusion principle.  The O-$p$ up-spin electron contributes in superexchange mechanism with Ru. If the magnetic moment of Ru is aligned anti-parallel to the moment of Mn, as shown in \fref{fig:lmro} (c), then O-$p$ up-spin electron can hop on both Ru-$e_g$ or Ru-$t_{2g}$ orbitals. However, in case of parallel alignment of magnetic moments of Mn and Ru, i.e., FM configuration, O-$p$ up-spin electron can only hop on Ru-$e_g$ orbitals, because Ru-$t_{2g}$ up-spin orbitals are almost fully occupied (see \tref{tab:n}). The Ru-$t_{2g}$ blockade decreases the kinetic energy gain in FM alignment and leads to an AFM ground state for LMRO (compare third $t_{2g}$ orbital occupation between AFM and FM configurations).  

\begin{table}
\begin{center}
    \begin{tabular}{c  c c }
    \hline
    \hline
     Ru-$d$ orbital&\multicolumn{2}{c}{Occupation $n_{\uparrow}$ ($n_{\downarrow}$)}\\ 
     &  AFM & FM   \\ \hline
         $t_{2g}$ & 0.60 (0.59)    &   0.62 (0.57) \\
         $t_{2g}$ &  0.55 (0.66)   &  0.65 (0.55) \\
         $t_{2g}$ &  0.22 (0.74)   &  0.74 (0.23) \\
         $e_{g}$  &  0.30 (0.46)   &   0.37 (0.42)   \\
         $e_{g}$  &  0.25 (0.30)   &    0.30 (0.26) \\
    \hline \hline
    \end{tabular}
    \caption{ Calculated (GGA+U) charge occupation of Ru-$d$ orbitals in La$_2$MnRuO$_6$ in AFM and FM magnetic configurations.  Numbers in parenthesizes denote $n_{\downarrow}$. Mn majority spin species is up. 
    }\label{tab:n}
\end{center}
\end{table}

\section{Electronic and magnetic properties of L\MakeLowercase{a}A''M\MakeLowercase{n}F\MakeLowercase{e}O$_6$ with A'' = B\MakeLowercase{a}, S\MakeLowercase{r} or C\MakeLowercase{a}}
\label{section:LAMFO}
In LMFO, strong electronic correlations in the Fe-$d$ orbitals favor Fe$^{3+}$ oxidation states in order to avoid double occupancy. This consequently leads to Mn$^{3+}$ states, Jahn-Teller distortion and, ultimately, a ferrimagnetic ground state. Here, we study LaA''MnFeO$_6$, with A''=Ba, Sr or Ca, in which the total oxidation of the cations at the B site is $7^+$ instead of $6^+$. This could lead to Mn$^{4+}$ and Fe$^{3+}$ oxidation states and to a ferromagnetic ground state. We investigate this possibility here.

\subsection{Structure optimization}
\label{section:LAMFO_struct}
Since doubly-ordered LA''MFO (A''$=$Ba, Sr, Ca) have not been reported experimentally, we start our study by optimizing their crystal structure. Most A$_2$B'B''O$_6$ double-perovskites with B-site order crystallize in the $P2_1/c$ space group.\cite{ref-a2bbo6} However, other space groups are also possible depending on the amount of octahedral distortion that is present in the crystal. One of the ways to predict the amount of octahedral tilting in a double-perovskite is from its tolerance factor $t$, which is defined by 
\begin{equation}
    t = \frac{\langle r_A \rangle + r_O}{\sqrt{2}(\langle r_B \rangle + r_O)},
    \label{eq:tolerance_factor}
\end{equation}
where $\langle r_A \rangle$ and $\langle r_B \rangle$ denote the average ionic radius at the A and B sites, respectively, while $r_O$ is the ionic radius of oxygen. 
The ideal, cubic situation with $180$\degree~B'-O-B'' bonding angles occurs when $t=1$. Most ordered double-perovskites with $t\simeq1$ crystallize in the $Fm\bar{3}m$ cubic space group. When $t$ is smaller than $1$, octahedral tilting occurs.\cite{ref-a2bbo6} The approximate tolerance factors of LBMFO, LSMFO and LCMFO calculated using \eref{eq:tolerance_factor} and available ionic radius values\cite{ref-rad} are $1.035$, $1.003$ and $0.985$ respectively. The decrease in tolerance factors is due to the fact that Ba$^{2+}$ has the largest ionic radius and Ca$^{2+}$ has the smallest one out of the three dopants.\cite{ref-rad} This motivates the study of the three following space groups for LSMFO and LBMFO: $P2_1/c$, $R\bar{3}$ and $Fm\bar{3}m$. In the case of LCMFO, its tolerance factor ($t=0.985$) is comparable to the tolerance factors of LMNO and LMCO ($t=0.978$ and $0.964$), both of which were found to crystallize in the $P2_1/c$ space group.\cite{ref-lmno, ref-lmco} For that reason, we only investigated this space group for this material.


In order to optimize the structures with unit cells containing a reasonable number of atoms, we had to impose A-site and B-site order. We chose a layered order on the A site and a rock-salt order on the B site, which are the most common orderings in doubly ordered A'A''B'B''O$_6$ double-perovskites.\cite{ref-aabbo6} 
We also tested rock-salt order on A and B site simultaneously for LBMFO and LSMFO. We found that the layered order on the A-site yields a lower total energy than the rock-salt A-site order.

For LBMFO, the ground state predicted by GGA+U calculations is in the $P2_1$ space group. We optimized the structures in the same space group for LCMFO and for the GGA calculations on both of these materials. 
In the case of LSMFO, the ground state predicted by GGA and GGA+U calculations is $P\bar{1}$. 
%
\begin{figure*}
    \begin{tabular}{ccc}
        \includegraphics[width=0.33\textwidth]{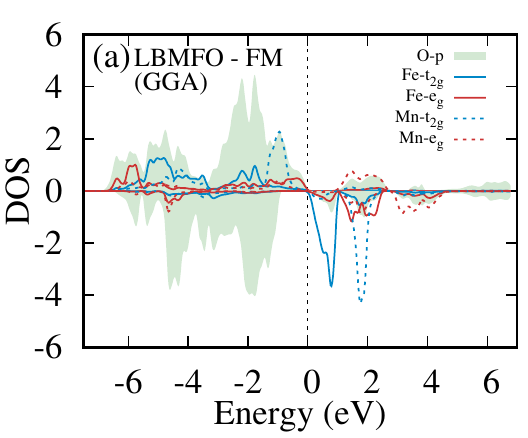}&
        \includegraphics[width=0.33\textwidth]{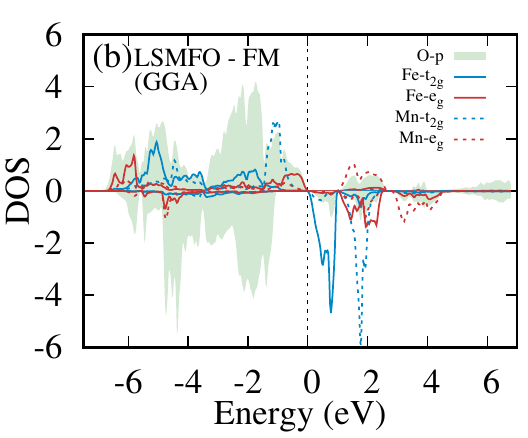}&
        \includegraphics[width=0.33\textwidth]{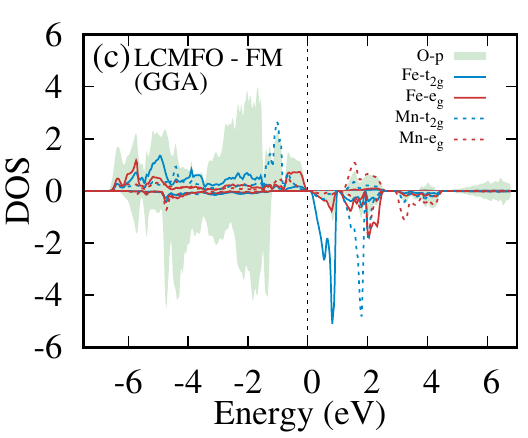}\\
        \includegraphics[width=0.33\textwidth]{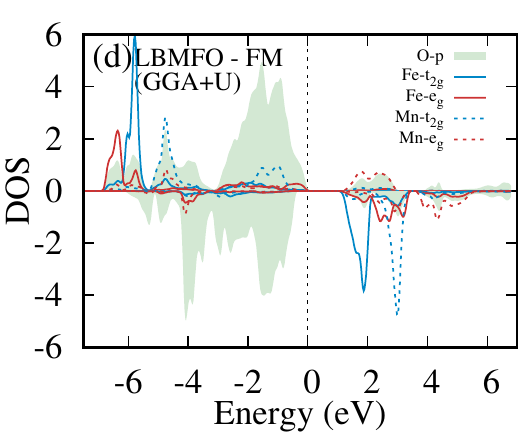}&
        \includegraphics[width=0.33\textwidth]{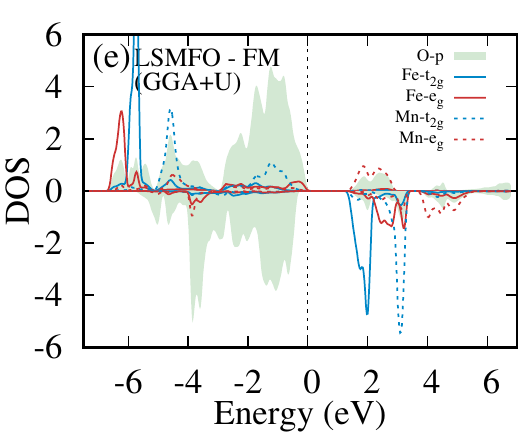}&
        \includegraphics[width=0.33\textwidth]{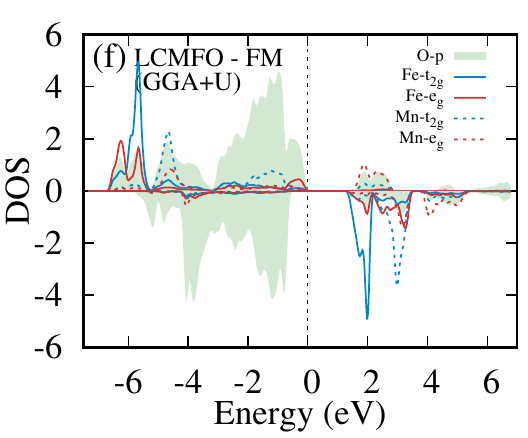}\\
        \\
        \multicolumn{3}{c}{\includegraphics[width=0.45\textwidth]{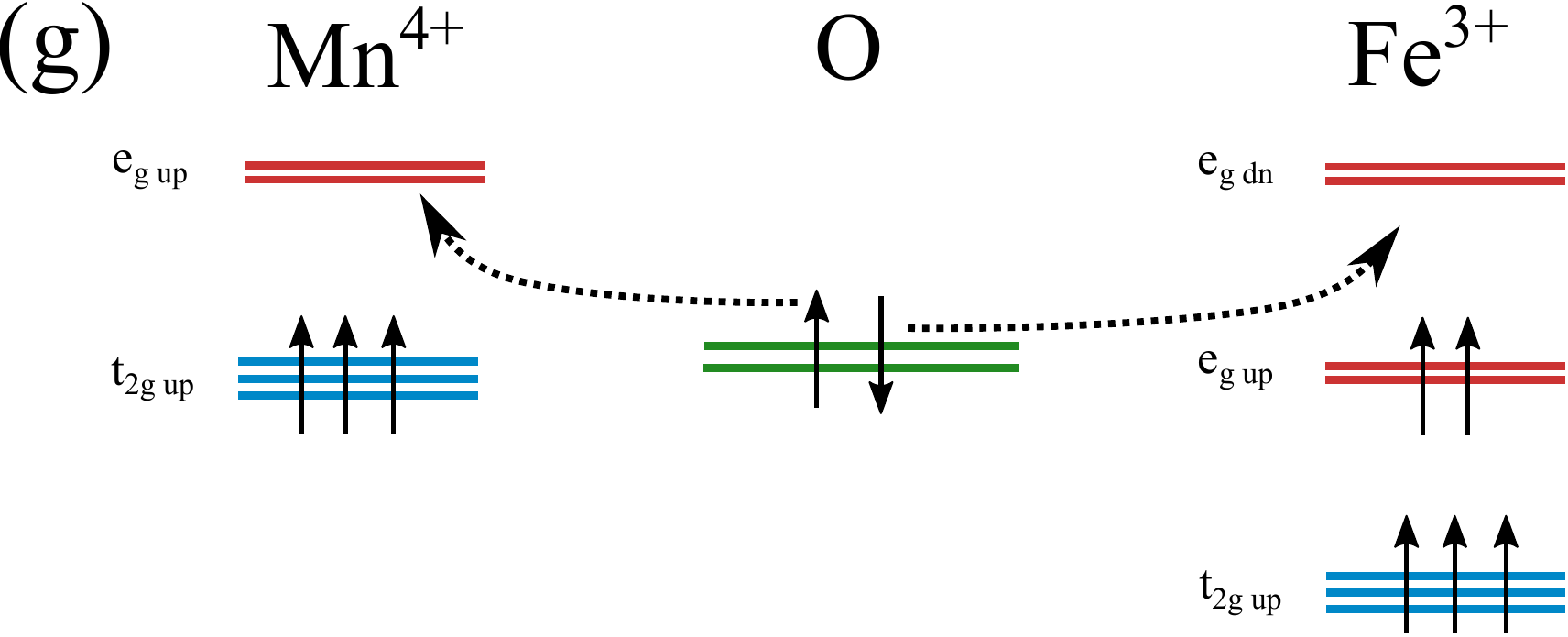}}  \\
        \end{tabular}
        \caption{(Color online) Top (middle) panel: Partial density of states for O-$p$, Fe-$e_g$ and Mn-$e_g$, and Fe-$t_{2g}$ and Mn-$t_{2g}$ orbitals from GGA (GGA+U) calculations. Positive DOS corresponds to the majority-spin channel, while negative DOS corresponds to the minority-spin channel. GGA and GGA+U calculations predict a ferromagnetic insulating ground state in (a, d) LBMFO, (b, e) LSMFO and (c, f) LCMFO.  Bottom panel (g): schematic representation of the ferromagnetic superexchange interaction between Mn$^{4+}$ and Fe$^{3+}$ in LBMFO, LSMFO and LCMFO.}
        \label{fig:lamfo}
    \end{figure*}

In agreement with the tolerance factors listed above, we found from the relaxed structures that the amount of octahedral tilting in LBMFO is the smallest out of the three compounds, while it is the largest in LCMFO. In previous work on double-perovskite LMFO ($t \simeq 0.96$), we found that the Mn-O-Fe bonding angles are $\simeq152$~\degree.\cite{ref-ourlmnopaper} The octahedral tilting in LMFO stems from the fact that La$^{3+}$ ions have a smaller ionic radius than the ideal case in which $t=1$. 
Since Ba$^{2+}$ and Sr$^{2+}$ ions have a larger ionic radius than La$^{3+}$, the Mn-O-Fe bonds in LBMFO and LSMFO are straightened to $\simeq172$\degree~and $\simeq165$\degree~respectively. Since the ionic radius of Ca$^{2+}$ is similar to that of La$^{3+}$, the Mn-O-Fe bonding angles are $\simeq157$\degree in LCMFO, comparable with LMFO.\cite{ref-rad}
\subsection{Magnetic and electronic ground state properties}
\label{section:LAMFO_dft}

All three compounds are predicted to be ferromagnetic insulators by both GGA and GGA+U methods. In the GGA calculations, the band gaps are very small for the three compounds. They are $0.14$, $0.17$ and $0.25$ eV for LBMFO, LSMFO and LCMFO respectively. As seen in \fref{fig:lamfo} (a), (b) and (c), the states that are immediately above the Fermi level are Fe-$t_{2g}$. Adding electron-electron interactions in the GGA+U calculations pushes the Fe-$t_{2g}$ states away from the Fermi level, and the predominant states above the Fermi level become Fe-$t_{2g}$ and Mn-$e_g$. The gaps open further to $\simeq 1.3$ eV in both LBMFO and LSMFO, and to $\simeq 1.4$ eV in LCMFO as seen in \fref{fig:lamfo} (d), (e) and (f). These figures also show that the occupied Fe-$e_g$ and $t_{2g}$ states are pushed to lower energies below the Fermi level. This localization is expected from the addition of electron correlations in these orbitals.  

The total moment is predicted to be $8\mu_B$/f.u. in all three materials. Analysis of the partial moment, partial density of states and partial charge in the five $d$ shells leads to the conclusion that the magnetic ions are in high spin configuration in all three ferromagnetic materials with Mn$^{4+}$ and Fe$^{3+}$ oxidation states. The magnetic orderings of these three compounds are easily understood from the superexchange interaction that can be deduced from these oxidation states. 

Indeed, as illustrated in \fref{fig:lamfo} (g), a ferromagnetic interaction between Mn$^{4+}$ and Fe$^{3+}$ is mediated through the oxygen $p$ electrons. Both O-$p$ electrons can hop on neighbouring $d$-shells when Fe and Mn $d$ electrons are ferromagnetically aligned, which results in an overall kinetic advantage. By contrast, in the antiferromagnetic case, the O-$p$ electron that has a spin aligned with those on Mn-$d$ orbitals can hop on the empty neighbouring Mn-$e_g$ orbital due to Hund's coupling, but the remaining O-$p$ electron cannot hop on Fe-$d$ shells due to Pauli's exclusion principle. This leads to a smaller kinetic advantage in the AFM case than in the FM case, which can explain why LBMFO, LSMFO and LCMFO are predicted to be ferromagnets. This intuition, supported by the Goodenough-Kanamori rules,\cite{PhysRev.100.564,KANAMORI195987} is confirmed by the partial DOS presented in \fref{fig:lamfo} (a) to (f). Here, Mn$^{4+}$ and Fe$^{3+}$ respectively have a $3d_{\sigma}^3d_{\bar{\sigma}}^0$ and a $3d_{\sigma}^5d_{\bar{\sigma}}^0$ electronic configuration, where $\sigma (\bar{\sigma})$ denotes the majority (minority) spin. For all materials, the partial DOS obtained from GGA and GGA+U calculations show a good overlap between O-$p_{\sigma}$ and Mn-$e_{g\sigma}$ above the Fermi level. There is also a good overlap between O-$p_{\bar{\sigma}}$ and Fe-$e_{g\bar{\sigma}}$, while the Mn-$e_{g\bar{\sigma}}$ partial DOS mainly lies at higher energies than both Fe-$e_{g\bar{\sigma}}$ and Mn-$e_{g\sigma}$.

All of these conclusions arise from the assumption that LBMFO, LSMFO and LCMFO are doubly-ordered. Experimentally, in A$_2$B'B''O$_6$ double-perovskites, B-site order seems to arise from charge and size difference between the B' and B'' ions. Typically, the materials are disordered when the charge difference is smaller than $2$.\cite{ref-a2bbo6} Moreover, A-site order seems to be linked with B-site order: if the B site is disordered, then the A site is also disordered.\cite{ref-aabbo6} These considerations indicate that doubly-ordered LBMFO, LSMFO and LCMFO could be difficult to synthesize experimentally, since the B-site charge difference is only of $1$. Disordered LA''MFO could include various domains, including Mn-O-Mn and Fe-O-Fe antiferromagnetic interactions. However, new experimental techniques seem to improve the degree of B-site order, which could then drive A-site order and lead to the ferromagnetic materials we describe here.\cite{Kleibeuker2017}

\subsection{Magnetic exchange couplings}
\label{section:LAMFO_exchange}
\begin{figure}
    \centering
\includegraphics[width=0.7\columnwidth]{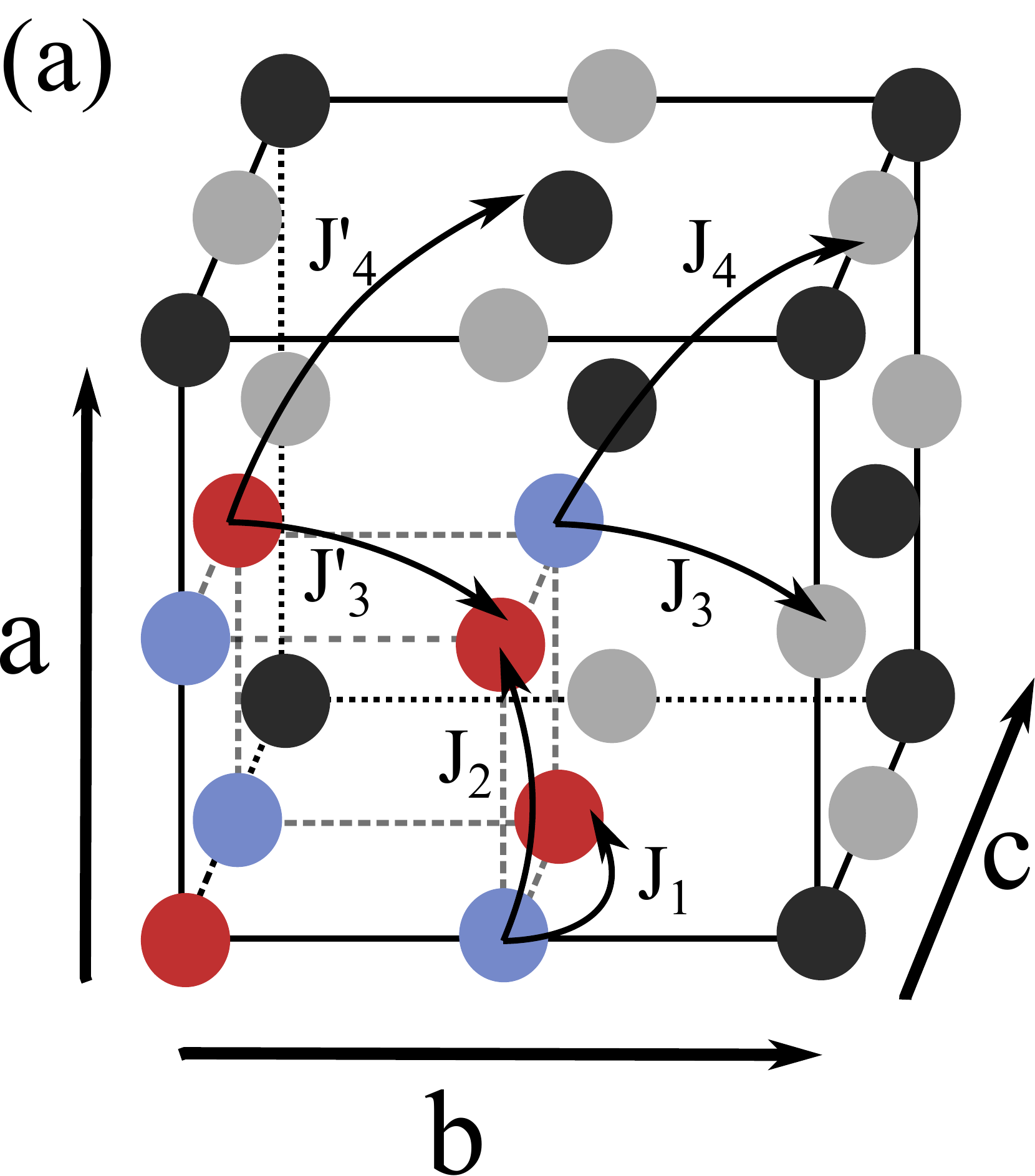}
    \caption{(Color online) Reduced supercell with the $4$ non-equivalent Mn (blue) and Fe (red) atoms. The lattice vector $\mathbf{a}$ denotes the out of plane direction, while the lattice vectors $\mathbf{b}$ and $\mathbf{c}$ generate the plane. }
    \label{fig:uncommon_afm}
\end{figure}

In order to see whether LBMFO, LSMFO and LCMFO are suitable for magnetic refrigeration, one needs to know if their $T_C$ is close to room-temperature. It is possible to map the DFT total energy to the Ising model 
\begin{equation}
    H = -\sum_{ij}J_{ij}S^z_iS^z_j,
    \label{eq:ising}
\end{equation}
in order to obtain the magnetic exchange couplings that can afterwards be used in the calculation of the Curie temperature. In \eref{eq:ising}, $J_{ij}$ denotes exchange coupling between magnetic moments at site $i$ and site $j$, while $S^z_{i(j)}$ is the $z$-component of the magnetic moment at site $i(j)$. We consider six independent exchange pathways connecting various Mn and Fe sites. We define $J_1$ and $J_2$ as the nearest-neighbor in-plane and out of plane couplings between Mn and Fe, while $J_3$($J'_3$) and $J_4$($J'_4$) are the next nearest-neighbor in-plane and out of plane couplings between Mn( Fe) magnetic moments. The exchange couplings are illustrated in \fref{fig:uncommon_afm} (a). 

In order to calculate the six exchange parameters, we fix the atomic positions and use seven different collinear magnetic configurations to calculate six total energy differences. We use an additional magnetic configuration to verify the validity of our results. All the magnetic configurations we consider are listed in \tref{tab:spin_config}. The configurations chosen here are the same as in Ref. \onlinecite{ref-ourlmnopaper}, where details of the mapping to the Ising model can be found. 
The exchange couplings are obtained from self-consistent GGA+U calculations on $\sqrt{2} \times \sqrt{2} \times 1$ supercells which include
$4$ non-equivalent Mn atoms, $4$ non-equivalent Fe atoms, and a total of $40$ atoms. 
\begin{table}
\begin{center}
    \begin{tabular}{ l  c  c  c }
    \hline
    \hline
    Configuration & Mn sublattice & Fe sublattice & Spin alignment \\ 
     \hline
    FM & i & i & in phase\\ 
    AFM1 (G-type) & i & i & out of phase \\ 
    AFM2 (A-type) & ii & ii & in phase \\ 
    AFM3 (C-type) & ii & ii & out of phase \\ 
    AFM4 & iii & iii & (see caption)  \\ 
    AFM5 & ii & iii & n.a.  \\ 
    FiM & i & ii & n.a. \\ 
    AFM6 & i & iii & n.a.  \\ 
    \hline
    \hline
    \end{tabular}
    \caption{Spin configuration of the sublattices used in the 8 magnetic configurations. The transition metal sublattice spin configurations are : (i) in-plane and out of plane FM (ii) in-plane FM and out of plane AFM and (iii) in-plane AFM and out of plane FM. For AFM4, the spin alignment of the sublattices is chosen in such a way that the out of plane nearest neighbor alignment is AFM. Configurations AFM5, AFM6 and FiM have different total energies, but relative spin alignment of the two sublattices in each of these configurations separately does not influence the expression of the total energy since there is no net contribution of nearest neighbor (Mn-Fe) interaction to the total energy. AFM6 is used to verify the validity of the results.
    }\label{tab:spin_config}
\end{center}
\end{table}
The values of the exchange couplings for LSMFO, LBMFO and LCMFO obtained from the mapping to the Ising model are listed in \tref{tab:js}. For all three materials, nearest neighbor exchange couplings $J_1$ and $J_2$ are ferromagnetic, as expected from the discussion on superexchange of the previous section. They are also larger by one or two orders of magnitude than the next-nearest neighbor couplings. This is due to the localization of the $3d$ orbitals. Moreover, most of the next-nearest exchange couplings are antiferromagnetic, as expected from the Goodenough-Kanamori rules regarding the superexchange interaction between two half-filled Fe-$d$ (Fe$^{3+}$-O-Fe$^{3+}$) or Mn-$t_{2g}$ (Mn$^{4+}$-O-Mn$^{4+}$) orbitals.\cite{PhysRev.100.564,KANAMORI195987} Finally, we computed the energy differences between the FM and AFM6 magnetic configurations using the calculated exchange couplings and compared them to the same energy difference obtained from GGA+U calculations. The energy differences are listed in \tref{tab:energy_afm6}. One can see that the agreement between the prediction from the calculated couplings and the DFT results is excellent for all three materials.

One can notice that LBMFO has the largest values of $J_1$ and $J_2$ LCMFO has the smallest ones. This trend can be explained by the structural differences between the three compounds. As explained in \sref{section:LAMFO_struct}, the Mn-O-Fe bonding angle is an interesting feature in the relaxed structures. Superexchange mechanisms are more effective when the two magnetic ions are aligned with the oxygen atom. A $180$\degree~bonding angle indeed leads to the biggest overlap between the orbitals that participate in the superexchange. In LBMFO, the angles are closer to the ideal $180$\degree~case, leading to more effective superexchange interactions. Similarly, in LCMFO, the angles are the farthest from $180$\degree: in that case, the crystal structure shows important distortions from the ideal double-perovskite one, which could lead to less effective superexchange interactions. Hence, the largest nearest-neighbor exchange couplings are obtained when the Mn-O-Fe bonding angle is close to $180$\degree. 
\begin{table}
\begin{center}
    \begin{tabular}{c  c c c  c }
    \hline
    \hline
     & Interaction path&\multicolumn{3}{c}{Values (meV)}\\ 
     &  & A''$=$Ba & Sr & Ca  \\ \hline
         $J_1$ & Mn-Fe (in plane) & 2.31 & 2.19  & 1.74 \\
         $J_2$ & Mn-Fe (out of plane) &  2.24  & 2.12 &  1.19 \\
         $J_3$ & Mn-Mn (in plane) &   -0.28  & -0.18 &  -0.12 \\
         $J_4$ & Mn-Mn (out of plane) &  -0.32  & -0.22  & -0.07   \\
         $J'_3$ & Fe-Fe (in plane)&  -0.003 
         & -0.02   &    0.03 \\
         $J'_4$ & Fe-Fe (out of plane) &  -0.01  & -0.03  &  -0.06 \\
    \hline \hline
    \end{tabular}
    \caption{ Calculated magnetic exchange interactions for LaA''MnFeO$_6$. Positive (negative) value denotes FM (AFM) coupling. The extreme values of the spins are $\mathcal{S}^z_{\rm Mn} = 3/2$ and $\mathcal{S}'^z_{\rm Fe} =5/2$. 
    }\label{tab:js}
\end{center}
\end{table}
\begin{table}
    \centering
    \begin{tabular}{c c c c}
    \hline
    \hline
         & \multicolumn{3}{c}{$E_{AFM6}-E_{FM}$} \\
         &  From $J$ values & & From ab initio calculations \\
         \hline
         A''$=$ Ba & 0.204 & & 0.205 \\
         Sr & 0.190 & & 0.191 \\
         Ca & 0.137 & & 0.138 \\
         \hline
         \hline
    \end{tabular}
    \caption{Energy difference between AFM6 phase and FM phase in eV.}
    \label{tab:energy_afm6}
\end{table}
\subsection{Curie temperature from mean-field and Monte Carlo calculations}
One can extract the Curie temperature of the materials using the exchange couplings listed in \tref{tab:js}. Here, we employ two different methods: a mean-field approximation and Monte Carlo calculations. The details on the mean-field and Monte Carlo calculations can be found in Ref. \onlinecite{ref-ourlmnopaper}. 
The resulting $T_C$s are listed in \tref{tab:temperatures}. The predicted $T_C$s are interesting from a qualitative point of view more than from a quantitative point of view. Indeed, without pretending that these values are accurate, we can still notice a trend in the predicted phase transition temperatures. Previous work using this methodology reproduced the qualitative experimental trend in $T_C$ for double-perovskites LMNO, LMFO and LMCO~\cite{ref-ourlmnopaper}. Calculations for LMFO using parameters similar to those we use here (GGA+U calculations with $U_{eff}=3$ eV, a supercell with $40$ atoms and the same magnetic orders) predicted its $T_N$ to be $329$ K (Monte Carlo) and $418$ K (mean-field). While the precise $T_N$ of LMFO is unknown, it was shown experimentally to be lower than the $T_C$ of both LMNO and LMCO, hence below room-temperature. 

This leads us to believe that, even though the predicted mean-field and Monte Carlo $T_C$s listed in table \ref{tab:temperatures} probably overestimate the actual $T_C$s, the trend that they follow should be accurate. We can notice that LBMFO has the highest $T_C$ while LCMFO has the lowest one for this family of compounds. This trend in $T_C$ can once again be explained by the structural differences between the three compounds. More importantly, by comparing our results for LBMFO, and LSMFO to the results obtained previously for LMNO in Ref.~\onlinecite{ref-ourlmnopaper}, we see that doubly-ordered LBMFO and LSMFO might have $T_C$s that are above or around room-temperature.

\begin{table}
    \centering
    \begin{tabular}{ccc}
    \hline
    \hline
          &  \multicolumn{2}{c}{Curie temperature (K)}\\
         & Mean-field & Monte-Carlo  \\
         \hline
         A''$=$Ba & 552 & 498 \\
         Sr & 530 & 425  \\
         Ca & 386 & 309  \\
         \hline
         \hline
    \end{tabular}
    \caption{Curie temperatures $T_C$ in Kelvin obtained from different methods for LA''MFO.}
    \label{tab:temperatures}
\end{table}

\section{Discussion: Magnetocaloric properties and potential for magnetic refrigeration}
In order to be adequate for magnetic refrigeration, a material must display a number of properties, among which are ferromagnetism with a $T_C$ around room temperature, as explained previously. For real applications, a material must also have an appreciable isothermal magnetic entropy change under the application of an external magnetic field $\Delta S_m(T,0\rightarrow H) = S_m(T, H)  - S_m(T,0) $ and a large adiabatic temperature change  $\Delta T_{ad}(S, H\rightarrow 0) = T_{ad}(S, H)  - T_{ad}(S, 0) $.

Here, we use mean-field calculations in order to compute the magnetic entropy of LBMFO and LSMFO. In order to check the accuracy of the method, we performed the same calculations for LMNO and compared our results to available experimental data. Appendix \ref{section:mf_mce} gives the details of the method. For LBMFO and LSMFO,  the extreme values of the spins are $\mathcal{S}^z_{\rm Mn} = 3/2$ and $\mathcal{S}'^z_{\rm Fe} =5/2$, and the exchange couplings are the ones from \tref{tab:js}. 

\begin{figure}
 \centering
\includegraphics[width=0.8\columnwidth]{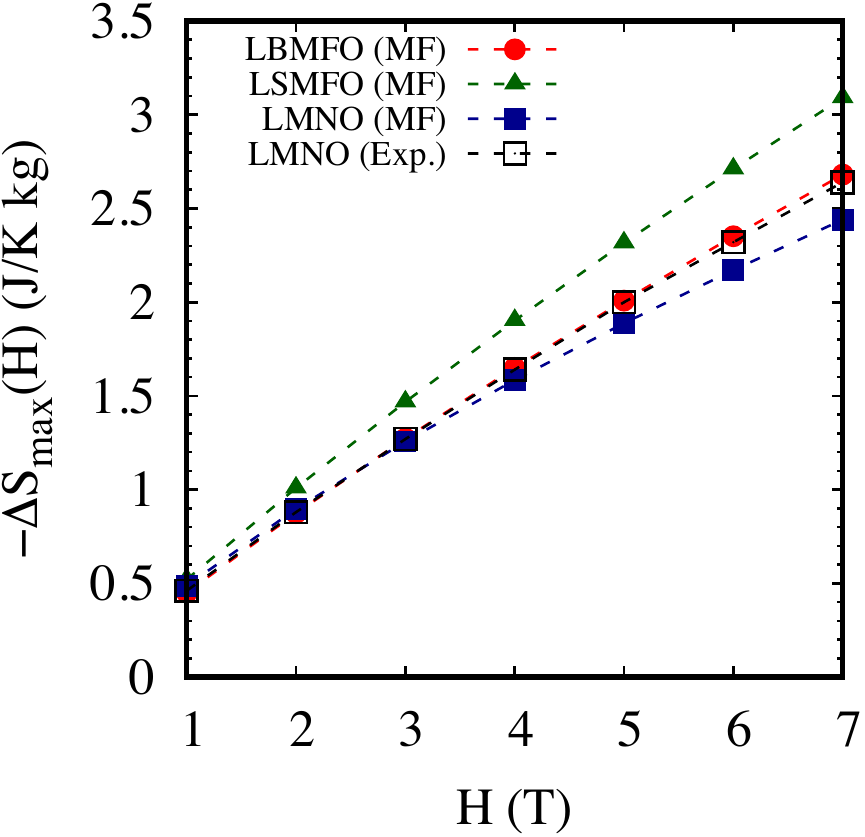}
    \caption{(Color online) Comparison between the experimental maximal value of the isothermal entropy change of LMNO\cite{ref-lmno} (black empty squares), and the mean-field calculated values of $\Delta S_{max} $ of LMNO (dark blue filled squares), LBMFO (red circles) and LSMFO (green triangles). We used for the number of formulas per kilogram, $N=1.25\times10^{24}$ and $N=1.40\times 10^{24}$ for LBMFO and LSMFO respectively to obtain the entropy change per kilogram of material from \eref{eq:entropy}.   }
    \label{fig:deltaSmax}
\end{figure}

 \fref{fig:deltaSmax}  shows the maximal value of the magnetic entropy change as a function of the external magnetic field for LBMFO, LSMFO and LMNO. For a field of $2$T, we obtain a value of $\sim1$J/K kg, which is about five times smaller than the maximal value of $\Delta S$ for reference material Gd.\cite{ref-gd} Similarly, we computed the maximal value of the adiabatic temperature change $\Delta T_{ad}$. The calculation of $\Delta T_{ad}$ requires the knowledge of the specific heat, which has not been reported for LSMFO and LBMFO in the literature. However, the specific heat of similar compound La$_2$MnCoO$_6$ has been measured.\cite{ref-lmco} It has been reported to fall rapidly below the Dulong-Petit limit around room temperature. Assuming a similar behaviour for LBMFO and LSMFO, we used the Dulong-Petit limit in order to compute a lower bound for $\Delta T_{ad}$. For an external field of $2$T, this yields a maximal value of $\Delta T_{ad}$ of about $0.2\%T_C$ for both LSMFO and LBMFO. In the mean field calculations, this is about $0.9$K, which is also a little over $5$ times smaller than in the case of Gd for the same field.\cite{ref-gd} 
 
It is a known fact that double perovskite oxides have smaller $\Delta S$ and $\Delta T_{ad}$ than other materials typically studied for their magnetocaloric properties. However, they offer a range of other properties that make them interesting for applications, such as their resistance to corrosion, their lower price and their high electric resistance.\cite{doi:10.1002/aenm.201200167} Therefore, even if their magnetocaloric properties are not to the level of that of reference materials, doubly-ordered LBMFO and LSMFO could still be promising candidates for magnetic refrigeration.

\section{Concluding remarks}\label{section:conclusion}
We studied two different types of materials in order to see if a ferromagnetic and insulating LaA''MnB''O$_6$-based double-perovskite with a higher moment per formula unit and Curie temperature than LMNO and LMCO could be designed. Since strong electronic correlations have been found to induce an antiferromagnetic insulating ground state in LMFO, we first studied LMRO, in which electronic correlations are less important. We found that this material is also predicted to be ferrimagnetic. The total moment is predicted to be 3$\mu_B$/f.u., which is higher than what was predicted with similar calculations for LMFO ($1\mu_B$). This predicted ground state and the fact that Ru is rather expensive strongly suggests that LMRO is not suitable for magnetic refrigeration.

By contrast, our study of hole-doped LMFO through divalent substitution of one of the La atoms leads to promising results. All three compounds studied (LBMFO, LSMFO and LCMFO) are predicted to be ferromagnetic insulators by GGA and GGA+U calculations, with a total moment of 8$\mu_B$/f.u, which is higher than that of LMCO and LMNO. Moreover, our study of the trend in $T_C$ 
of these hole-doped LMFO materials indicate that LBMFO and LSMFO are likely to have a $T_C$ close to room-temperature, making both of them promising for room-temperature magnetic refrigeration, provided that they can be synthesized experimentally as doubly ordered doped double-perovskites. Even though maximum isothermal entropy change and adiabatic temperature change with magnetic field are smaller than for the best reference materials, these double perovskites have other advantages, such as high resistance, low price and chemical stability, that can make them useful for applications.

\begin{acknowledgments}
We are indebted to T.E. Baker who performed the Monte Carlo calculations. We are grateful to M.~Balli, P.~Fournier and M. C\^ot\'e for useful discussions. This work has been supported by the Natural Sciences and Engineering Research Council of Canada (NSERC) under grant RGPIN-2014-04584, by the Canada First Research Excellence Fund, and by the Research Chair in the Theory of Quantum Materials. Ab initio simulations and Monte Carlo calculations were performed on computers provided by the Canadian Foundation for Innovation, the Minist\`ere de l'\'Education des Loisirs et du Sport (Qu\'ebec), Calcul Qu\'ebec, and Compute Canada.
\end{acknowledgments}
\appendix
\section{Mean-field calculation of the magnetocaloric properties}
\label{section:mf_mce}

We used the two-spin mean-field approach to the Ising model described in Ref. \onlinecite{ref-ourlmnopaper}. This mean-field scheme takes into account the two magnetic ions of LaA''MnB''O$_6$ (A'' = La, Ba or Sr, B'' = Ni or Fe). It yields the following self-consistent equations for the magnetizations $m(T)$ and $m'(T)$ per Mn and B'' site, considering an external magnetic field $H$:
\[
\begin{aligned}
m(T) &= \mathcal{B}(h_{MF}+ g\mu_B H, \mathcal{S}^z),\\
m'(T) &= \mathcal{B}(h'_{MF}+ g\mu_B H, \mathcal{S}'^z),
\end{aligned}
\]
where the mean fields $h_{MF}$ and $h'_{MF}$ are given by:
\[
\begin{aligned}
h_{MF} &= 2(4J_1+2J_2)m'+2(4J_3+8J_4)m,\\
h'_{MF} &= 2(4J_1+2J_2)m+2(4J'_3+8J'_4)m',
\end{aligned}
\]
and $ \mathcal{B}(h, \mathcal{S}^z)$ is a Brillouin function:
\[
\begin{aligned}
\mathcal{B}(h, \mathcal{S}^z)&=-\frac{1}{2}\coth(\beta h/2) \nonumber\\
&+(\mathcal{S}^z+\frac{1}{2})\coth(\beta h(\mathcal{S}^z+1/2)).
\end{aligned}
\]
Here, $\mathcal{S}^z $ is the maximal spin value at the Mn site,  $\mathcal{S}'^z$ is the maximal spin value at the B'' site. Also, $\beta = 1/k_BT$,  the $J$s are the exchange-coupling constants discussed in \sref{section:LAMFO_exchange}. Self-consistent solution of these equation for a given temperature gives the sublattice magnetization and the corresponding mean fields.

The magnetic entropy $S_m(T,H)$ in the presence of an external magnetic field $H$ is defined as:
\[
\begin{aligned}
S_m(T,H) &= -\frac{\partial F_m}{\partial T}  \nonumber \\
&= Nk_B\left ( \ln{(Z)} + T \frac{1}{Z}\frac{\partial Z}{\partial T} \right ) \nonumber \\
&+ Nk_B\left ( \ln{(Z')} + T \frac{1}{Z'}\frac{\partial Z'}{\partial T} \right ) \nonumber\\
&= S^{Mn}_m(T,H) + S^{B''}_m(T,H),
\end{aligned}
\]
where $F_m$ is the free energy, $N$ the number of formulas in the sample per kilogram and the sublattice partition functions are
\[
\begin{aligned}
 Z
&=\sum_{S^z_i} \exp[\beta (h_{MF}+g\mu_B H) S^z_i],\\
Z'
&=\sum_{S'^z_i} \exp[\beta (h'_{MF}+g\mu_B H) S'^z_i].
\end{aligned}
\]

The final equation for the entropy due to the Mn spins is:
\begin{align}\label{eq:entropy}
S^{Mn}_m(T,H)& = Nk_B \ln \left [ \frac{\sinh{(\beta(h_{MF}+g\mu_B H)(\mathcal{S}^z+1/2)})}{\sinh{(\beta(h_{MF}+g\mu_B H)/2})}\right ]\nonumber\\ 
&- N \frac{(h_{MF}+g\mu_B H)}{T}\mathcal{B}(h_{MF}+g\mu_B H, \mathcal{S}^z).
\end{align}
The expression for the entropy due to the B'' magnetic atom is analogous.

We computed $h_{MF}$ and $h'_{MF}$ for LMNO from the self-consistency equations, and then calculated the entropy from \eref {eq:entropy}, with $\mathcal{S}^z = 3/2$ and  $\mathcal{S}'^z = 1$. In particular, we calculated the isothermal entropy change $\Delta S_m(T,0\rightarrow H) = S_m(T, H)  - S_m(T,0) $ for LMNO. This mean field theory can also be used to compute a lower limit for the adiabatic temperature change $\Delta T_{ad}$, using the Dulong-Petit limit for the heat capacity: $C_p = 3nR$, where $n$ is the number of atoms in a unit cell, and $R$ the universal gas constant. We use the approximate form of the $\Delta T_{ad}$:
\begin{equation}\label{eq:deltaTad}
\Delta T_{ad}(T, H) = \frac{T}{C_p} \Delta S_m(T, 0\rightarrow H)
\end{equation}

\begin{figure}
 \centering
\includegraphics[width=0.7\columnwidth]{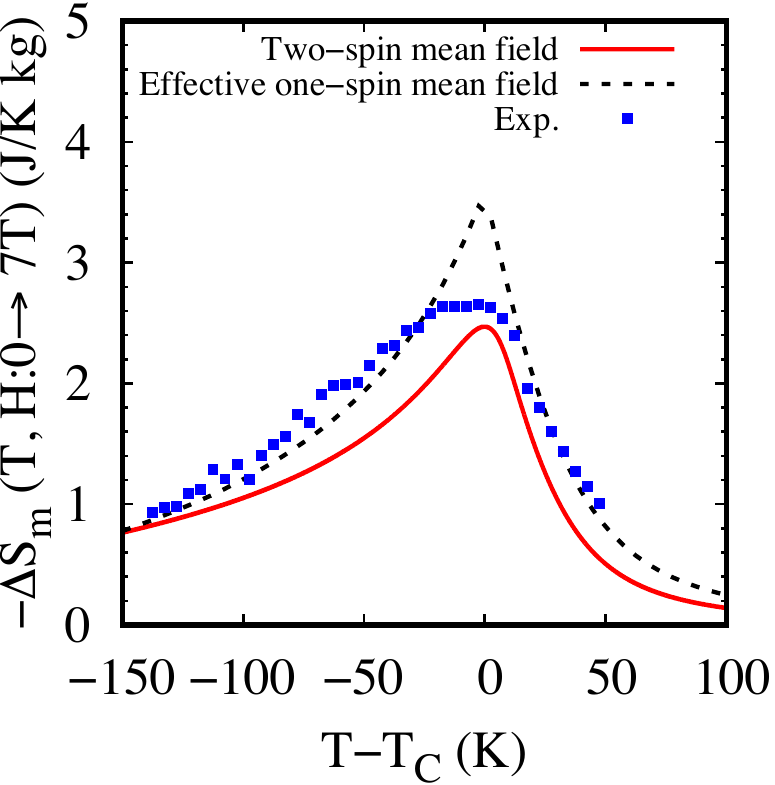}
    \caption{(Color online) Isothermal entropy change for an external field of $7$T in LMNO from our two-spin mean-field calculation (full red line), and from experimental (blue dots) and effective one-spin mean-field (dashed black line) data.\cite{ref-lmno}  We plotted $\Delta S$ as a function of $T-T_C$, where $T_C = 280$K for the experimental and simple mean-field data, and $T_C=527$K for our mean-field results.} 
    \label{fig:deltaS_LMNO}
\end{figure}

\fref{fig:deltaS_LMNO} compares $\Delta S_m(T-T_C,0\rightarrow 7~\text{T})$ from our two-spin mean-field calculation with the experimental and mean-field data from Ref. \onlinecite{ref-lmno}. The mean-field model of Ref. \onlinecite{ref-lmno} is an effective one-spin model where the effective angular momentum $J$ of each unit cell ($J=2.75$) is coupled to its first neighbours. In our calculations, we used $N=1.235\times10^{24}$ formula units, which corresponds to one kilogram of LMNO. The behavior of the calculated $\Delta S_m$ around $T_C$ agrees qualitatively with the experiment, but the full width at half maximum is smaller. 

One can also notice that the maximal value $\Delta S_{max}$  is better reproduced by the present two-spin mean-field approach than by the effective one-spin mean-field approach of Ref. \onlinecite{ref-lmno}. This is illustrated in \fref{fig:deltaSmax}, which shows the evolution of $\Delta S_{max}$ with the external magnetic field. Our mean-field results reproduce the experimental data for LMNO quite well, especially at low external field. This relatively good agreement between the experimental data and our mean-field results leads us to believe that the mean-field approach can be a good tool to characterize the magnetocaloric effect in LMNO and, by extension, to gain insight on the magnetocaloric effect in LBMFO and LSMFO.

%

\end{document}